# Colloidal-quantum-dot spasers and plasmonic amplifiers

Stephan J. P. Kress[1]*, Jian Cui[1]*, Patrik Rohner[2], David K. Kim[1], Felipe V. Antolinez[1], Karl-Augustin Zaininger[1], Sriharsha V. Jayanti[1], Patrizia Richner[2], Kevin M. McPeak[1], Dimos Poulikakos[2], and David J. Norris[1]

[1]Optical Materials Engineering Laboratory, ETH Zurich, 8092 Zurich, Switzerland.
[2]Laboratory of Thermodynamics in Emerging Technologies, ETH Zurich, 8092 Zurich, Switzerland.

*These authors contributed equally to this work.

**Colloidal quantum dots[1] are robust, efficient, and tunable emitters now used in lighting[2-6], displays[7], and lasers[8-11]. Consequently, when the spaser[12]—a laser-like source of surface plasmons—was first proposed[13], quantum dots were specified as the ideal plasmonic gain medium. Subsequent spaser designs[14-17], however, have required a single material to simultaneously provide gain and define the plasmonic cavity, an approach ill-suited to quantum dots and other colloidal nanomaterials. Here we develop a more open architecture that decouples the gain medium from the cavity, leading to a versatile class of quantum-dot-based spasers that allow controlled generation, extraction, and manipulation of plasmons. We first create high-quality-factor, aberration-corrected, Ag plasmonic cavities. We then incorporate quantum dots via electrohydrodynamic printing[18,19] or drop-casting. Photoexcitation under ambient conditions generates monochromatic plasmons above threshold. This signal is extracted, directed through an integrated amplifier, and focused at a nearby nanoscale tip, generating intense electromagnetic fields. This spaser platform, deployable at different wavelengths, size scales, and geometries, can enable more complex on-chip plasmonic devices.**

Plasmons are surface-bound electromagnetic waves that propagate along metal interfaces[20]. If their motion is confined in a cavity between two plasmonic reflectors, only standing waves (or modes) at specific wavelengths are permitted. By adding a gain medium that provides energy at these wavelengths, plasmonic losses can be compensated[21]. Specifically, propagating plasmons can stimulate the emission of additional plasmons, introducing amplification. When the gain exceeds losses for a specific cavity mode, the plasmonic analog of a laser, or spaser, results. An intense narrowband plasmonic signal is produced that can then be extracted for use. For example, by routing this signal towards a sharp tip, monochromatic plasmons can be concentrated in nanoscale volumes for enhancing light–matter interactions.

In the original spaser proposal[13], both the generation and concentration of plasmons occurred within the same nanoscale cavity—a metallic particle coated with colloidal quantum dots. Quantum dots[1] were specified for gain because of their high emission quantum yields (often >90%), large transition dipole moments, good photostability, and ability to be packed densely without suffering the self-quenching effects observed in dyes. However, while the proposed nanoparticle spaser can potentially provide intense local fields, it requires extremely high gain[22] and does not produce propagating plasmons. Thus, many efforts have aimed at directly amplifying plasmons propagating within larger devices that exhibit lower losses[14-17].

These works have followed two general approaches. Top-down nanofabrication has been used to define a semiconductor region that is encased in metal[15,23], a design that so far has not permitted easy extraction of the cavity plasmons. Alternatively, bottom-up methods have been used to place single-crystalline semiconductor nanowires[14,17] or nanoflakes[16] on a dielectric-coated metal surface. The volume between the metal, the nanostructure, and its end facets define the plasmonic cavity. However, this approach, which exploits the high gain of single-crystalline semiconductors to overcome the modest reflectivity at their facets, has not been amenable to a broader class of useful colloidal nanomaterials. Notably, colloidal quantum-dot spasers have not been realized despite the long use of these materials in lasers.[8-11] In addition, because the nanowires or nanoflakes are randomly positioned on a surface, they are difficult to integrate with other elements on a chip.

To address these issues, we have applied a hybrid strategy in which we separately engineer the plasmonic cavity and the gain medium. We first optimized the cavity by defining and positioning reflectors for propagating plasmons on a metal surface via lithography. In addition to high reflectivity and placement accuracy, this results in an open and adaptable cavity design that allows quantum dots (or another colloidal nanomaterial) to be easily added for gain. Our strategy also focused on the versatility of the structure, while compromising on its dimensions. When extreme confinement of the generated plasmons is desired, our spaser signal can be concentrated outside the cavity through an additional integrated element, as shown below.

Our cavities consist of two 400- to 600-nm-tall Ag block reflectors protruding from an ultrasmooth Ag surface (Fig. 1a). These blocks provide the in-plane plasmon reflectivity[24,25] (>90%) necessary to obtain high-quality-factor ($Q$) cavities. To produce these structures, we first created the inverse of the desired pattern on a [100]-oriented Si wafer with electron-beam lithography and reactive-ion etching (see Section S1 in the Supplementary Information). Following thermal evaporation of Ag[26] onto this Si template, we peeled off the Ag film with epoxy via template stripping[27]. By reusing the Si template, many copies of the same chip could be obtained. We then placed CdSe/CdS/ZnS core/shell/shell quantum dots[25,28,29] directly onto the planar Ag surface between the reflectors using electrohydrodynamic NanoDrip printing[18,19] (Fig. 1b; see Sections S2-S3 in the Supplementary Information). With the combined capabilities of electron-beam lithography and NanoDrip printing, devices can be designed and positioned at will (Fig. 1c).

In conventional laser cavities, stable modes are supported between aberration-corrected concave mirrors[30]. Analogously, our plasmonic reflectors were designed with an aspherical parabolic correction to produce cavities with stable, well-defined plasmonic modes (see Sections S4-S6 in the Supplementary Information). We focused on cavities with reflectors 10 µm apart with a radius-of-curvature at their apex twice their spacing (R=2L cavity). The spatial profile of the resulting mode, calculated using a simple ray-tracing and reflection algorithm, is highlighted in red in Fig. 2a.

To spectrally characterize the modes of our cavities when empty, we printed ~100 quantum dots at their center. Photoexcitation of the quantum dots launches plasmons into cavity modes[19]. Some of these plasmons scatter into photons at the outer edge of the block reflector. Thus, by measuring the far-field scattering spectrum, the cavity modes within the quantum-dot emission bandwidth can be revealed. Figure 2b compares the measured (red curve) and predicted (grey curve) modes for a 10-µm cavity (see Sections S7-S9 in the Supplementary Information). The slight offset between the two is consistent with a slightly smaller experimental cavity (9.965 µm). The measured linewidths were typically 3 nm (9 meV). For a perfect structure, we predicted linewidths of 2.2 nm (6.8 meV) when we included the finite reflectivity of the blocks and the measured permittivity for our Ag.

Spatially, calculations suggest that these modes have a ~1-µm beam waist (Fig. 2a) and are confined within ~100 nm of the surface when the quantum-dot layer is present (see Figs. S1 and S2 in the Supplementary Information). Thus, to ensure spatial overlap and maximize gain for spasing, we printed quantum-dot stripes that were 2 µm wide and at least 100 nm thick (Fig. 2c and Fig. S3). Low-intensity photoexcitation then yielded cavity spectra as in Fig.

2d. Due to losses from the quantum dots, particularly absorption, the modes broaden and only ripples appear on the long-wavelength side. The spectral mode separation also decreases due to the higher refractive index *n* in the cavity because of the quantum-dot layer (Fig. 2d, grey curve).

Under pulsed excitation, the cavity spectra change significantly with increasing power density. At 130 µJ/cm$^2$ (Fig. 2e), a small feature appears near the middle of the spectrum at a plasmonic mode position. At 250 µJ/cm$^2$ (Fig. 2f), this feature becomes much sharper and more intense. The insets in Fig. 2e,f show real-space images (false color) of the same cavities. At 250 µJ/cm$^2$, photon emission from the cavity center greatly decreases while scattering from the reflectors increases. These changes in both the spectral and spatial distribution of scattered plasmons with increasing excitation are consistent with stimulated emission and amplification of plasmons into a single mode defined by the cavity, *i.e.*, the onset of spasing.

Quantum dots allow the wavelength of this effect to be easily tuned (Fig. 3a). We fabricated devices with 602-, 625-, and 633-nm-emitting quantum dots. Some devices (see 633) exhibited single-mode operation with typical linewidths of 2 meV, or 0.65 nm, indicating a *Q* approaching 1000 (Fig. 3b). Other devices (see 602 and 625) showed multiple modes that changed in relative intensity with increasing excitation. Such mode competition, common in lasers[30], should also occur in spasers[31].

Figure 3c shows the output power spectrally integrated over the peak of the single-mode 633 device as a function of excitation power. A noticeable inflection occurs at ~180 µJ/cm$^2$ that corresponds to the spasing threshold. (Other devices had values down to 100 µJ/cm$^2$.) This is somewhat surprising given that the best quantum-dot lasers, which should have lower losses, exhibit comparable values[10]. While this certainly attests to the quality of our devices, other factors must be involved. Thresholds can be reduced in spasers due to the Purcell enhancement of the quantum-dot emission rate and high coupling of this emission to the cavity mode (see Sections S10-S12 and Figs. S4-S5 in the Supplementary Information for further discussion).

Green-emitting quantum dots did not exhibit spasing, presumably due to increased plasmonic losses at shorter wavelengths. However, at very high excitation intensities (~1000 µJ/cm$^2$), new spectral features appeared for our red-emitting quantum dots (Fig. 3d). Under these conditions, higher-energy multiexcitons (emitting around 590 nm) are generated in the quantum dots, providing sufficient gain[32] for spasing into the green (578 nm). Moreover, a progression of spasing modes was observed with spacings in agreement with calculations assuming *n*=1.60 for the quantum-dot film (Fig. 3d, dotted curve). While this refractive index is smaller than that measured via ellipsometry (1.75), such high exciton populations can significantly decrease *n*[33]. A lower *n* also agrees with observed blue-shifts in the spasing modes with increased excitation.

Having demonstrated spasing, we now extract this signal for potential on-chip use. Our device architecture allows for straightforward integration of a plasmonic waveguide simply by extending the reflector. Moreover, as the plasmons propagate out of the cavity along the waveguide, they can be focused to the nanoscale by tapering[34]. Figure 4a shows a cavity in which one reflector has been extended ~35 µm and narrowed to a tip. Figure 4b presents the same device operating above threshold. The contrast has been enhanced (10$^4$) in the right-hand portion of the image to resolve plasmons that are guided and focused. The cavity spectrum (Fig. 4c) matches that of plasmons scattered at the tip (Fig. 4d). Thus, the spaser signal is extracted, guided, and focused, but with diminished intensity due to propagation losses.

The tip signal can be further enhanced if the spaser is combined with a plasmonic amplifier. While this could be achieved by printing quantum dots on the waveguide, we took advantage of the solution-processability of colloidal nanomaterials and developed a simple drop-casting method that conformally coats our entire chip with a ~150-nm-thick film of quantum dots. Figure 4e,f shows this approach applied to an extended-reflector cavity (14° taper). Below threshold (Fig. 4e), broadband emission is observed that is fairly uniform with some guiding and focusing of plasmons at the tip. Above threshold (Fig. 4f), the spatial intensity distribution completely changes. The output of the spaser is amplified as it travels towards the tip, leading to highly concentrated plasmons[34]. The emission from the tip in Fig. 4h with amplification is more than $10^3$ times stronger than in Fig. 4d without. When comparing the spectrum from the cavity (Fig. 4g) with that from the tip (Fig. 4h), we observe the same modes. However, at the tip, spontaneous emission is diminished and a single spaser mode is selectively amplified. At higher excitation intensities, unwanted amplified spontaneous emission can occur from the amplifier alone (see Fig. S6 in the Supplementary Information). However, this leads to many finely spaced features, unrelated to the cavity modes, in the tip spectrum.

Our measurements were performed on chips containing arrays of devices. These operated under air at room temperature with near-unity yield (108 out of 112, or 96% of our measured drop-casted devices showed spasing). While we have presented individual functional units consisting of a spaser, amplifier, and waveguide, many such units can be linked to build an integrated circuit. Going beyond block reflectors, elements such as plasmonic antennas and gratings could be implemented within the current scheme. Quantum dots (or other gain media such as organic-inorganic perovskites, rare-earth-doped oxides, *etc*.) can then be deposited where needed for spasing and signal amplification. The flexibility and broad applicability of this platform can enable the practical implementation of spasers for integrated plasmonic applications.

**Acknowledgements**

We thank E. De Leo, S. Meyer, Y. Fedoryshyn, and U. Drechsler for experimental assistance. We gratefully acknowledge funding from the European Research Council under the European Union's Seventh Framework Program (FP/2007-2013) / ERC Grant Agreement Nr. 339905 (QuaDoPS Advanced Grant) and from the Swiss National Science Foundation under Grant Nr. 146180. J.C. acknowledges funding from the ETH Zurich Postdoctoral Fellowship Program and the Marie Curie Actions for People COFUND Program.

**Author contributions**

S.J.P.K., J.C., and D.J.N. conceived the ideas and planned the experiments. S.J.P.K. designed and fabricated the resonators with assistance from J.C. and K.M.M. D.K.K. synthesized the quantum dots. P.Ro. performed NanoDrip printing with assistance from P.Ri. and supervision from D.P. K-A.Z. developed the dropcasting method. J.C. performed optical measurements with assistance from S.J.P.K. and S.V.J. S.J.P.K., J.C., P.Ro., and F.V.A. analyzed the data with guidance from D.J.N. S.J.P.K., J.C., and D.J.N. wrote the manuscript with input from all authors.

**Additional information**

Supplementary information including methods is available in the online version of the paper. Reprints and permissions information is available online at www.nature.com/reprints. Correspondence and requests for materials should be addressed to D.J.N. (dnorris@ethz.ch).

**Competing financial interests**

D.P. is involved with a start-up company that is attempting to commercialize NanoDrip printing. Otherwise, the authors declare no competing financial interests. Readers are welcome to comment on the online version of the paper.

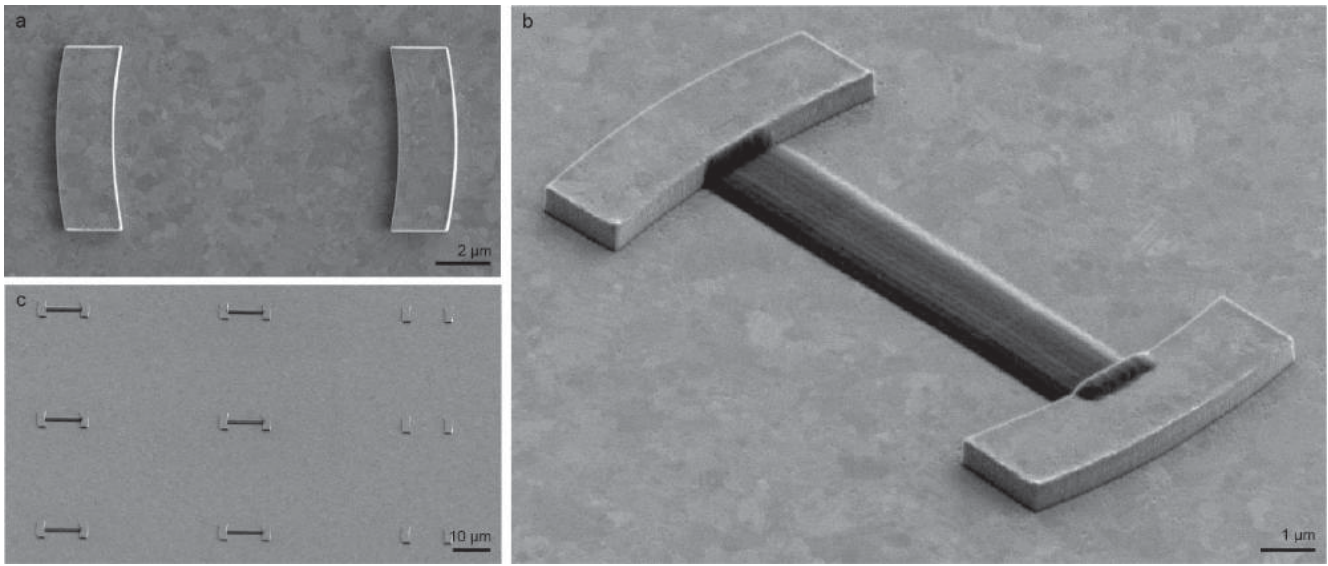

**Figure 1 | Structure of colloidal-quantum-dot spaser. a**, Top-view electron micrograph of a plasmonic cavity: two ~600-nm-tall Ag block reflectors positioned 10 µm apart on an ultrasmooth Ag surface. The blocks can be designed and placed at will. Plasmons propagate between the reflectors to create a cavity. **b**, Tilted view of a functional device. NanoDrip printing is used to deposit a stripe (~100 nm thick and ~2 µm wide) of colloidal quantum dots between the reflectors. **c**, Tilted view of a square array of 9 plasmonic cavities (50 µm apart). The three cavities in the right column are empty. The rest contain a quantum-dot stripe, as in **b**.

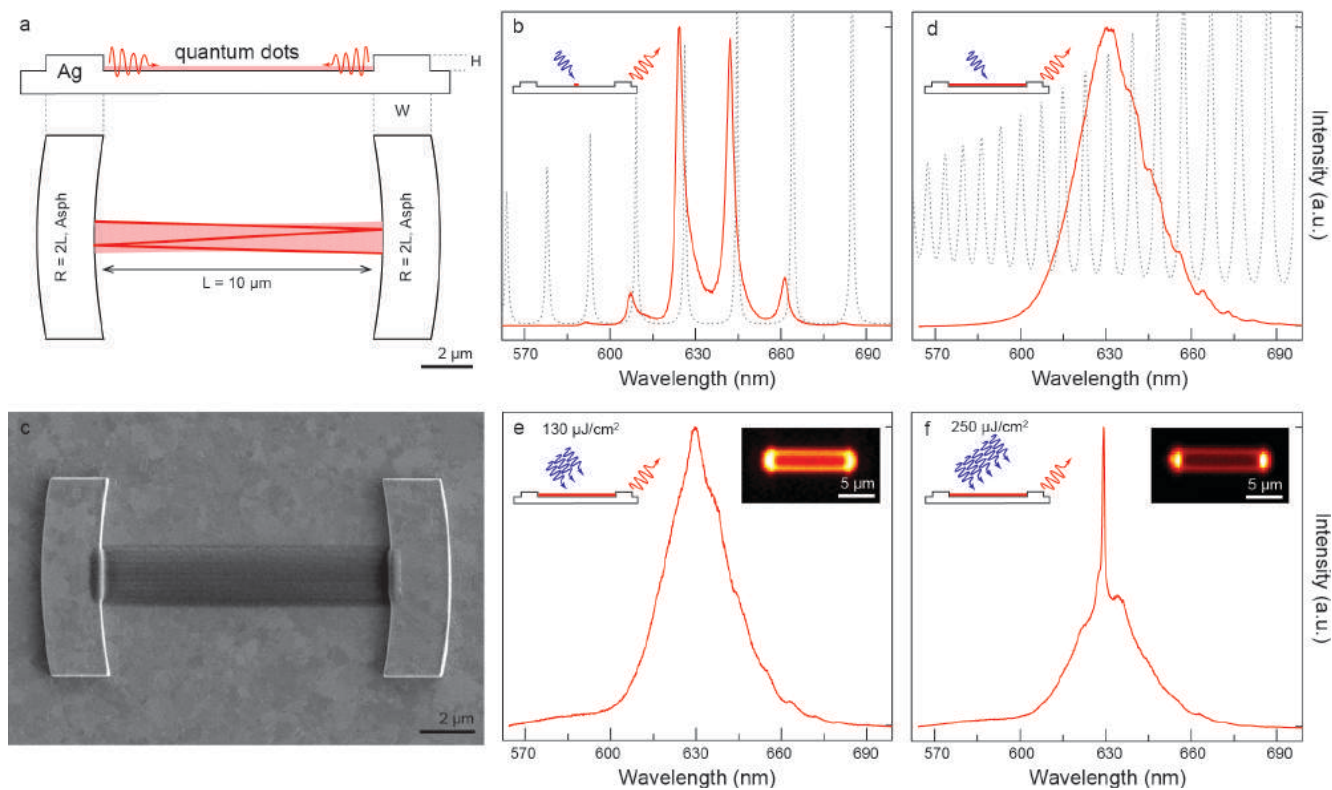

**Figure 2 | Design and characterization of plasmonic cavities containing quantum dots. a**, The plasmonic reflectors are designed with a radius-of-curvature twice the cavity length (R=2L cavity) and include an aspherical parabolic correction. Plasmons emitted into the cavity are spatially confined in a stable cavity mode, depicted in red. **b**, When ~100 quantum dots are placed in the cavity and weakly photoexcited, they emit into plasmonic cavity modes. The experimental cavity spectrum, measured by detecting plasmons scattered into the far field at the outer edge of one of the reflectors, is shown in red. The calculated spectra of the stable plasmonic cavity modes are in grey (see Sections S8-S9 in Supplementary Information). The slight offset is due to a slightly smaller experimental cavity length (9.965 µm). **c**, Top-view electron micrograph of the cavity for comparison with **a**. The quantum-dot stripe is printed to maximize spatial overlap with the cavity mode. **d**, When a cavity filled with quantum dots as in **c** is weakly photoexcited, modes appear as ripples in the cavity spectrum (collected as in **b**) only at longer wavelengths due to losses from the quantum-dot film. Calculated modes (neglecting quantum-dot absorption) are in grey. **e**, Under pulsed excitation (130 µJ/cm$^2$) a small feature rises on top of the cavity spectrum at the position of a calculated plasmonic mode. Right inset: the real-space image (false color) of the emission from the quantum-dot stripe. The two bright spots are due to scattering off the reflectors. **f**, At higher excitation (250 µJ/cm$^2$) the small peak in **e** narrows and increases in intensity. Right inset: real-space image as in **e**. The device exhibits decreased emission within the stripe and increased signal at the reflectors. The changes in the spectra and images in **e**,**f** are indicative of the onset of spasing. Cartoons in **b** and **d**-**f** depict the optical excitation and collection processes.

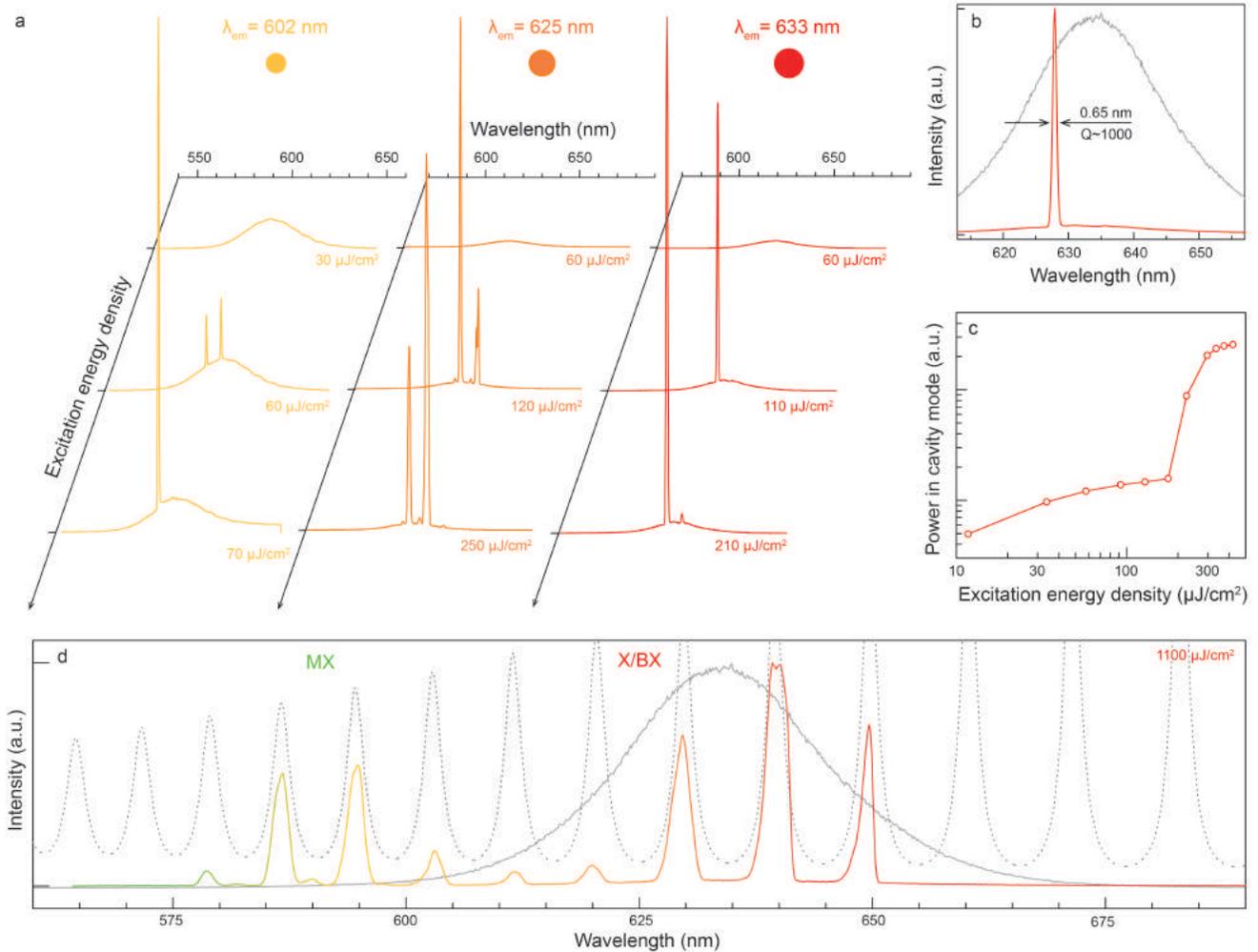

**Figure 3 | Analysis of quantum-dot spasers. a**, Quantum-dot samples with emission centered at 602, 625, and 633 nm were used to fabricate spasers. Cavity spectra for each are plotted for three excitation intensities, one below and two above threshold. **b**, The 633 device from **a** exhibited a single spasing mode above threshold (red curve). Typical linewidths were 2 meV (0.65 nm, Q ≈ 1000). The broad photoluminescence spectrum from quantum dots on flat Ag (outside the cavity) is shown for comparison (grey curve). **c**, The input-output power plot of the 633 device (red points) reveals an inflection at ~180 μJ/cm$^2$. Spasing thresholds as low as ~100 μJ/cm$^2$ were observed. **d**, When the 633 device is excited at very high intensities (~1000 μJ/cm$^2$), a progression of spasing modes is observed. These occur within the gain envelopes of the exciton/biexciton (X/BX) centered at ~640 nm and the multiexcitons (MX) centered at ~590 nm. The ~160 meV spacing between the envelopes is consistent with prior measurement (Ref. 32). The broad quantum-dot spectrum (solid grey curve) from **b** and the expected plasmonic modes for a film with *n* = 1.60 (dashed grey curve) are shown for comparison.

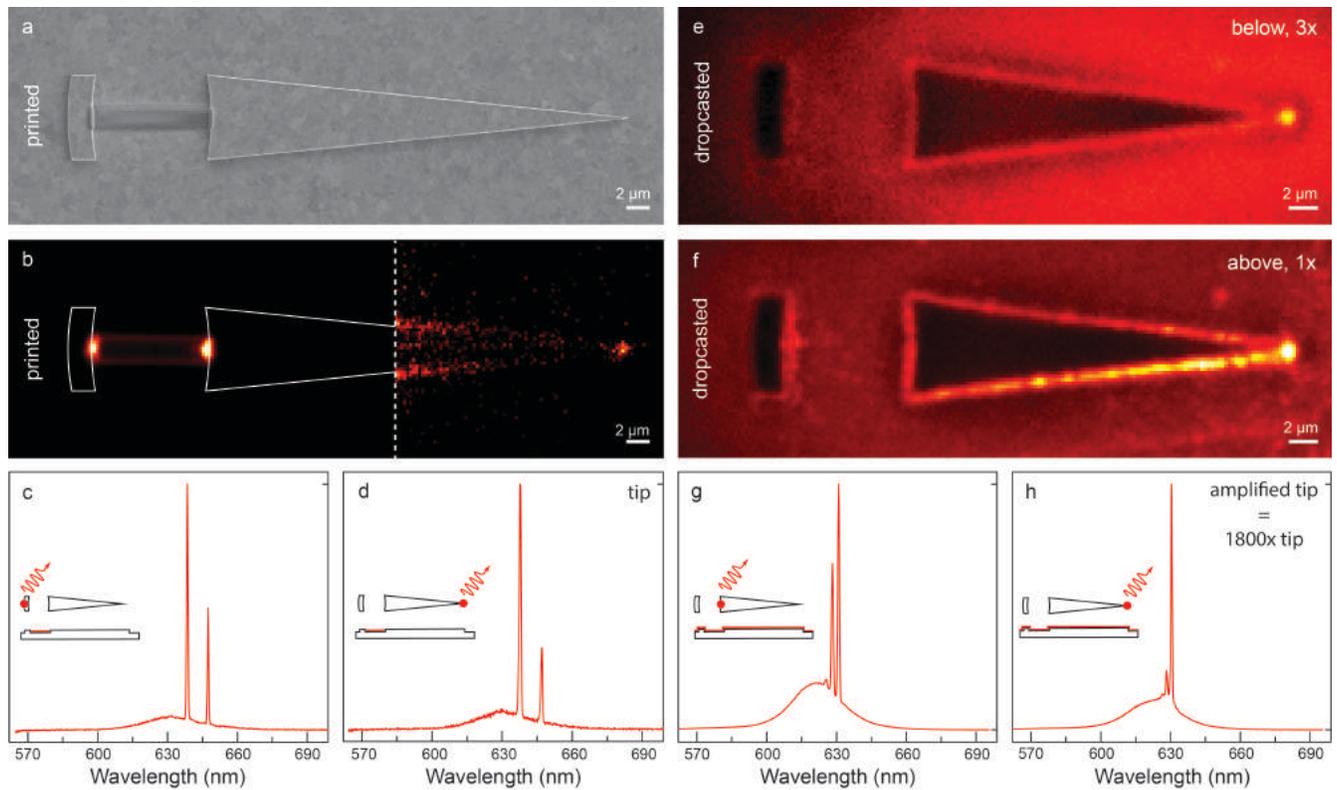

**Figure 4 | Extraction, amplification, and focusing of plasmons generated by quantum-dot spasers. a**, Top-view electron micrograph of a spaser with a printed quantum-dot stripe and an elongated reflector (11.5° taper) to guide and nanofocus the spaser signal. **b**, Real-space image (false color) of the device in **a** above threshold. The contrast of the image to the right of the vertical dashed line is enhanced ($10^4$) to show the plasmons focused at the tip. **c,d**, Spectra measured by collecting plasmons scattered at the outer reflector edge (**c**) and the tip (**d**) confirm that the spasing signal is guided and focused. **e,f** Real-space images (false color) of a cavity with an elongated reflector (14° taper) coated with a ~150-nm-thick drop-casted quantum-dot film, excited below and above threshold, respectively. The intensity of **e**, which is multiplied by 3, is fairly uniform across the device with some broadband focusing at the tip. In **f**, plasmons generated by the cavity are amplified and focused at the tip. **g,h**, Spectra measured by collecting plasmons scattered at the inner reflector edge (**g**) and the tip (**h**), confirm that a single spasing mode is selectively guided, amplified, and focused. The signal in **h** (with amplification) is 1800 times greater than in **d** (without).

# Colloidal-quantum-dot spasers and plasmonic amplifiers

Stephan J. P. Kress[1]*, Jian Cui[1]*, Patrik Rohner[2], David K. Kim[1], Felipe V. Antolinez[1], Karl-Augustin Zaininger[1], Sriharsha V. Jayanti[1], Patrizia Richner[2], Kevin M. McPeak[1], Dimos Poulikakos[2], and David J. Norris[1]

[1]Optical Materials Engineering Laboratory, ETH Zurich, 8092 Zurich, Switzerland.
[2]Laboratory of Thermodynamics in Emerging Technologies, ETH Zurich, 8092 Zurich, Switzerland.

*These authors contributed equally to this work.

## Supplementary information

**S1. Plasmonic-cavity fabrication.** Two-inch Si(001) wafers with a native-oxide layer were used as the template. After a pre-bake step (180 °C, 10 min), ~280 nm of an electron-beam resist (Allresist, CSAR 62 AR-P 6200) was spin-cast onto each wafer (2000 r.p.m., 60 s). This was followed by a post-bake step (150 °C, 5 min). The resist was exposed using an electron-beam lithography system (Vistec, EBPG5200, acceleration voltage 100 kV, aperture size 300 µm, and base dose 290 µC/cm$^2$). Following exposure, the resist was developed (Allresist, AR 600-546), and the Si was etched using an inductively coupled plasma deep reactive-ion etch (Oxford, Plasmalab System 100, HBr, 40 sccm, 80 W RF). Afterwards, the remaining resist was removed using oxygen plasma (100 W, 5 min) and exposure to NMP (Sigma-Aldrich, n-methyl-2-pyrrolidone, >60 min at 130 °C). Following a rinse with isopropanol, the Si templates were loaded into a thermal evaporator (Kurt J. Lesker, Nano36) containing a home-built rotating sample holder tilted 30° from horizontal. ~1 µm of Ag (Kurt J. Lesker, 99.99%) was deposited at 50 Å/s under a chamber gas pressure of <10$^{-7}$ Torr while the sample rotated at ~60 r.p.m. After evaporation, the plasmonic cavities were manually template-stripped from the Si using a glass microscope slide bonded to the surface with a ultraviolet-light-curable epoxy (Epoxy Technology, OG142-95)[S1,S2]. The direction of template stripping was chosen to minimize structural defects at the reflector walls.

**S2. Quantum-dot synthesis.** Core/shell/shell CdSe/CdS/ZnS colloidal quantum dots were synthesized according to previously published recipes[S2]. Their color was tuned by using smaller CdSe cores while adjusting the shells accordingly.

**S3. Localized quantum-dot printing.** Quantum dots were transferred from hexane to tetradecane through selective evaporation. The concentration was adjusted to an absorbance of 0.5 at the lowest energy exciton absorption feature for a cuvette with a 1-mm path length. These inks were further diluted 1:3, 1:2, and 1:5 in tetradecane for the 602-, 625-, and 633-nm-emitting quantum dots, respectively. A description of the electrohydrodynamic NanoDrip printing setup can be found elsewhere[S3]. Additionally, a 50x objective (Edmund Optics, Plan Apo ULWD, NA 0.42, working distance 20.5 mm) was mounted at an angle of 45° for live inspection. Printing nozzles were fabricated by pulling glass capillaries (TW100-4, World Precision Instruments) using a Sutter Instruments P-97 pipette puller. These were then coated (electron-beam evaporation, Plassys, MEB550S) with a 10-nm titanium adhesion layer and a 100-nm layer of gold. The outer diameter of the nozzles was in the range of 1800 to 2000 nm. A DC electric potential of ~250 V was applied between the nozzle (+) and the template-stripped silver substrate (ground). The separation between the nozzle tip and the substrate was 5 µm. The 10-µm long, 2-µm wide and ~100-400-nm-thick quantum-dot stripes (*e.g.*, see Fig. 2c) were deposited by moving the substrate with a velocity of 6 µm/s in a serpentine-like and layer-by-layer fashion with a line pitch of 160 nm. 7 to 16 layers were added to obtain the desired thicknesses. Atomic force microscope (AFM) topographical images (see Fig. S3) were measured in tapping mode with a Bruker Dimension FastScan AFM with a TESPA-v2 tip (8 nm tip radius).

**S4. Computation of mode stability.** To find stable plasmonic cavity geometries, we exploit the conditions derived for laser-cavity design via the matrix-transfer method. For symmetric cavities that have two mirrors of identical curvature, a single geometric parameter, $g = 1 - \frac{L}{R}$, determines the stability. Here, $L$ is the inter-mirror spacing and $R$ is the mirror curvature. The matrix-transfer method shows that a symmetric cavity supports stable, reproducing modes when $-1 < g < 1$. Throughout this work, we used a symmetric cavity with, $g = 0.5$, or mirrors that have a radius-of-curvature $R$ that is twice the inter-mirror-spacing, $R = 2L$.

**S5. Beam-waist calculation and ray tracing.** To better understand the propagation and extent of the mode in our cavity, we performed beam waist and ray tracing calculations. Beam waist calculations were computed assuming that the modes in our cavities are well-approximated by 2D-Gaussian modes. 2D ray-tracing between the reflectors was performed using a simple in-house MATLAB code (Fig. 2a in the main text).

**S6. Cavity design.** The symmetric cavities in this work consist of two mirrors with an inter-mirror spacing of L = 10 µm. These mirrors have a concave, parabolic geometry with an apex-radius-of-curvature that is two times the inter-mirror-spacing (R = 2L = 20 µm). The reflector blocks were typically designed to be 400 to 600 nm in height (H) to enable high reflectivity of the incident plasmons. The blocks had a width, W (see Fig. 2a), of 2 µm to avoid any detrimental shadowing effects during silver evaporation.

**S7. Optical characterization: Optical pumping, imaging, and spectroscopy.** For weak optical excitation (Fig. 2b,d), a white-light light-emitting diode (Lumencor, Sola SE II light engine) was used in conjunction with a dichroic beamsplitter (488 nm long pass, AHF Analysentechnik). The remaining images and spectra were obtained using a pulsed source. 450-nm pulses (~340 fs pulse duration, 1 KHz repetition rate) were generated by a collinear optical parametric amplifier (Spectra-Physics, Spirit-OPA) pumped by a 1040 nm pump laser (Spectra-Physics, Spirit-1040-8). After spectral filtering, the beam was directed through a gradient neutral density filter wheel to adjust the pulse power (Thorlabs, NDC-50C-2M-B). Following beam expansion and collimation, a small portion of the beam was directed to a photodiode to monitor the pump power (Thorlabs, DET110). The remainder of the beam was passed through a defocusing lens (focal length of 150 mm) into an inverted microscope (Nikon, Eclipse Ti-U). The beam was then directed upwards to the sample using a dichroic beamsplitter (488 nm long pass, AHF Analysentechnik) through a 50x air objective (Nikon, TU Plan Fluor, 0.8 NA).

The defocusing lens was adjusted to provide a spot size of 60 to 70 μm. The spot size was determined from an image of the photoluminescence from a flat portion of a film of quantum dots. A cross-section of the spot was then fitted with the sum of two Gaussian functions and the full-width-at-half-maximum was determined numerically. The excitation power density at the sample was monitored by correlating the power meter reading above the objective at the sample plane (Thorlabs, S170C with PM100D) with the electrical current reading from the photodiode.

Emission from the sample was collected by the same objective and directed through the dichroic beamsplitter (500-nm-long-pass emission filter, AHF Analysentechnik) and relay lenses (focal length of 200 mm) into an imaging spectrometer (Andor, Shamrock 303i). The emission was dispersed with a 300 lines/mm grating (500 nm blaze) and imaged with an air-cooled electron-multiplying CCD camera (Andor, iXon 888 Ultra). Real-space images were obtained using the zero-order mode of the same grating.

**S8. Plasmonic Fabry-Pérot calculation for cavities.** The modes of our planar plasmonic cavities can be modeled using a lossy one-dimensional Fabry-Pérot resonator,

$$\frac{I_{res}}{I_0} = \frac{1}{(1-r_{cav})^2 + 4\,r_{cav}\sin^2\left(\frac{\pi\nu}{\nu_{cav}}\right)},$$

where $\frac{I_{res}}{I_0}$ gives the intensity modulation with frequency, $\nu$, induced by the presence of the reflectors. The spectrum (*i.e.* strength and spacing) of these modes depends on the round-trip loss factor for the electric field,

$$r_{cav} = R \cdot e^{-l_0/L_{mode}},$$

and the characteristic cavity frequency,

$$\nu_{cav} = \frac{c_0}{2 n_{mode} l},$$

where $R$ is the plasmon reflectivity (intensity) of the reflectors, $c_0$ is speed of light in vacuum, and $l_0$ is the geometric size of the cavity. $L_{mode}$ and $n_{mode}$ are the propagation length and effective index of the mode considered.

To determine the effective cavity length $l$, the geometric cavity length $l_0$ has to be corrected by the penetration depth into the silver, $\delta$ (obtained from the Fresnel equations)[S4],

$$l = l_0 + 2\delta.$$

In particular, $L_{mode}$ and $n_{mode}$ are strongly dispersive (as expected for plasmons), whereas the penetration depth, $\delta$, and reflection coefficient, $R$, are only weakly dependent on the frequency $\nu$. Both $L_{\text{mode}}$ and $n_{\text{mode}}$ are also dependent on the quantum-dot-layer thickness, $t_{\text{QD}}$, and refractive index, $n_{\text{QD}}$, as well as the dielectric function of the silver films, $\varepsilon_{\text{Ag}}(\nu)$.

**S9. Calculation of propagating modes in silver/quantum-dot/air stack.** To obtain values for the mode index, $n_{\text{mode}}(n_{\text{QD}}, t_{\text{QD}}, \varepsilon_{\text{Ag}}, \nu)$, and propagation length, $L_{\text{mode}}(n_{\text{QD}}, t_{\text{QD}}, \varepsilon_{\text{Ag}}, \nu)$, for this infinite three-layer system (silver/quantum-dot-layer/air), we numerically determined these from analytical expressions as previously reported[S5]. This analysis allows us to retrieve all modes (both photonic and plasmonic) propagating along the interface in our three-layer system. Once we solve for the complex wave-vectors, $k_{\text{mode}}(n_{\text{QD}}, t_{\text{QD}}, \varepsilon_{\text{Ag}}, \nu)$ for the propagating modes are determined, and we can deduce the effective index,

$$n_{\text{mode}} = \Re\left(\frac{k_{\text{mode}}}{k_0}\right),$$

and propagation length,

$$L_{\text{mode}} = \frac{1}{\Im(2k_{\text{mode}})},$$

of the modes, which are functions (see Figs. S1 and S2). For the plasmonic mode in an infinitely thick quantum-dot-layer, simple expressions for the effective refractive index,

$$n_{SPP}(n_{\text{QD}}, \varepsilon_{\text{Ag}}, \nu) = \Re\left\{\sqrt{\frac{n_{\text{QD}}^2 \varepsilon_{\text{Ag}}(\nu)}{n_{\text{QD}}^2 + \varepsilon_{\text{Ag}}(\nu)}}\right\},$$

and propagation length of the plasmonic mode,

$$L_{SPP}(n_{\text{QD}}, \varepsilon_{\text{Ag}}, \nu) = \frac{c_0}{4\pi\nu} \Im\left\{\sqrt{\frac{n_{\text{QD}}^2 + \varepsilon_{\text{Ag}}(\nu)}{n_{\text{QD}}^2 \varepsilon_{\text{Ag}}(\nu)}}\right\},$$

can be found.

**S10. Optical characterization: Lifetime measurements.** Time-resolved measurements were performed on the same microscope setup as described above in Section S7. The pump laser was adjusted to 100 kHz for these measurements with power reduced to <1 nJ/cm$^2$. The emission was focused onto an avalanche photodiode (Picoquant, MPD-SPAD) carefully placed at the image plane of a different exit port of the same spectrometer. This allowed us to spatially resolve the emission from the sample by moving the sample position using a piezo nanopositioning system (Mad City Labs, Nano-LPS300). A time-correlated single-photon-counting module (Picoquant, PicoHarp 300) was used to acquire time-resolved data. The time delay histograms were binned to 64 ps bin-widths and fit to a biexponential curve for the first 50 ns of the decay. These data were scaled such that the fit was normalized at t = 0 (Fig. S4). Different measurements were compared at the 1/*e* value of the decay. The emission was collected from photoluminescence of quantum dots printed on fused silica, quantum dots printed on flat Ag, quantum dots printed into one of our cavities, and from plasmon scattering from the outer edge of the block reflector of the same cavity.

**S11. Spasing thresholds.** We have observed spasing thresholds as low as 100 µJ/cm$^2$ in our devices. As mentioned in the main text, this is somewhat surprising given that the best quantum-dot lasers, which should have lower losses, exhibit comparable values. The decreased thresholds may be attributed to Purcell enhancement of the quantum-dot emission rate, high coupling of this emission to the cavity mode, and high modal overlap. Time-resolved measurements suggest that the Purcell enhancement in our devices could be as high as ~2 (Fig. S4). However, this estimate does not account for the effect of possible quenching of the emitter by the metal. The increase in output power above threshold (Fig. 3c), which, according to traditional laser rate equations[S6], scales inversely with the effective coupling factor β, indicates a relatively high effective value of ~0.06. Finally, the overlap between the plasmonic TM0 and the quantum-dot layer is very high (see Fig. S1). These factors contribute to lower spasing thresholds as propagating plasmons can be more efficiently generated and stimulated.

**S12. Modal gain, plasmonic loss, and quantum-dot material gain.** A mode is amplified when the gain provided by the gain medium exceeds loss. The modal gain, $g_{mode}$, is given by the following expression,

$$g_{mode} = \Gamma \cdot g_{QD} - \alpha_{mode} .$$

Here, losses in the mode are included in $\alpha_{mode}$, the modal loss coefficient (loss per unit length). The total gain is given by $\Gamma \cdot g_{QD}$, where, $g_{QD}$, is the material gain provided by the quantum-dot layer, and the modal overlap factor $\Gamma$ quantifies the spatial overlap of the mode and the gain medium. The modal overlap factor for the surface-plasmon-polariton mode is close to unity, $\Gamma \sim 1$, due to strong confinement (Fig. S1). This is also evident from the high effective index of the mode (Fig. S2). Since reflection losses are negligible compared to propagation losses (computed from the measured permittivity for our silver obtained from ellipsometry), the latter determine the modal losses and we calculate a loss of $\alpha_{mode} = 810 \text{ cm}^{-1}$. With $\Gamma \sim 1$, we thus determine a minimum material gain needed for amplification of $g_{QD} = 810 \text{ cm}^{-1}$. This is close to values reported[S7] for photonic devices with quantum dots synthesized using a similar recipe ($g_{QD} = 750 \text{ cm}^{-1}$). However, for our system, we measured the modal gain on the tapered amplifier (as in Fig. 4 in the main text) to be $g_{mode} = 500 \text{ cm}^{-1}$ (Fig. S5), which implies a higher material gain from the quantum dots ($g_{QD} = 1310 \text{ cm}^{-1}$). This increase in gain may be attributed to enhancements in the Purcell factor (up to ~2) and $β$- coupling factor in our plasmonic device compared to the photonic case, as discussed in Section S11 above.

## Supplementary references

# Supplementary figures

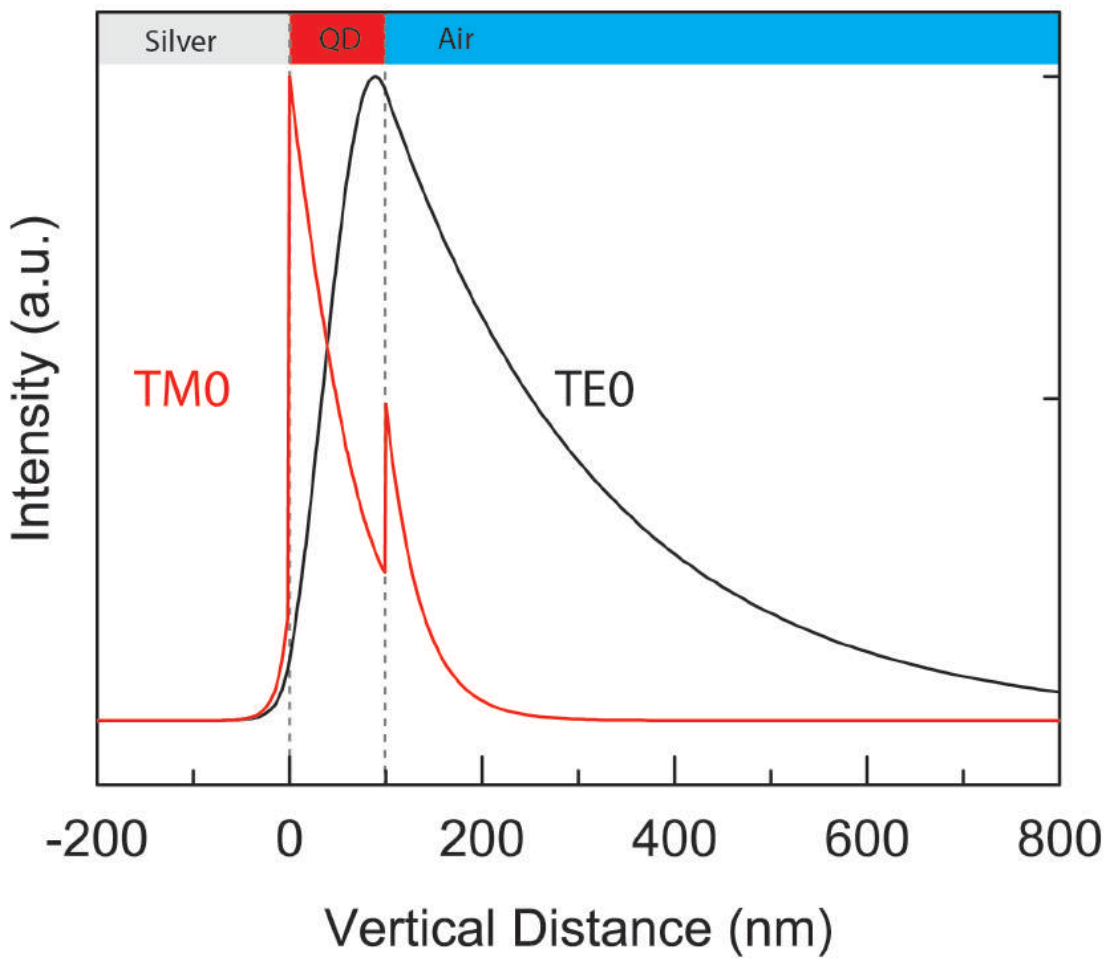

**Supplementary Figure S1 | Electric-field profiles.** The absolute value squared of the electric fields of the fundamental transverse magnetic (TM0) and fundamental transverse electric (TE0) modes calculated for a 100-nm-thick quantum-dot film on flat Ag. The plasmonic TM0 mode is confined closely to the Ag surface with high spatial overlap with the quantum dots. The photonic TE0 layer is extended far above the Ag surface and therefore has poor modal overlap with the quantum dots and decreased reflectivity from the block reflectors, which are 400 to 600 nm tall.

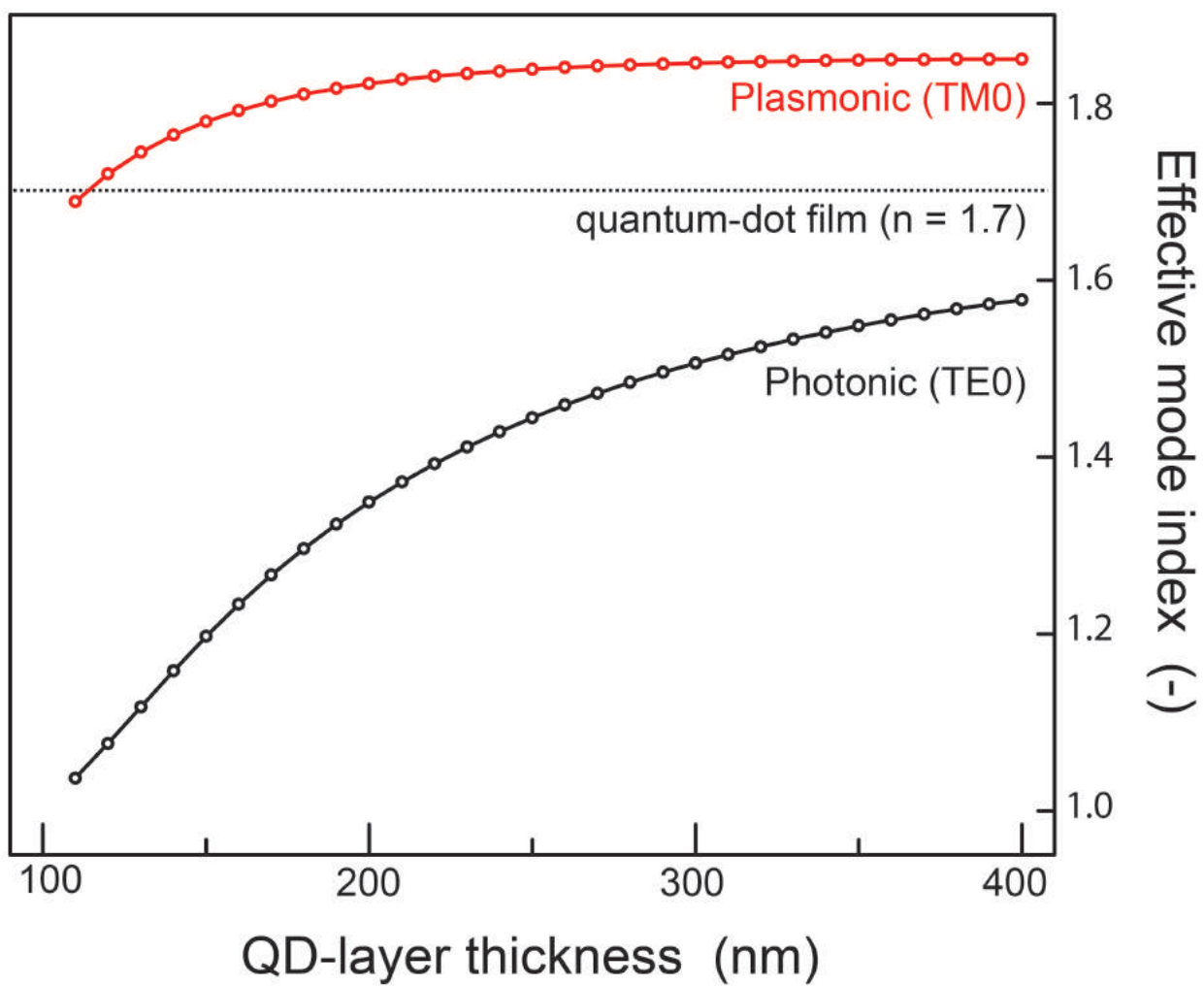

**Supplementary Figure S2 | Calculated effective indices for plasmonic and photonic modes.** The effective indices of refraction are plotted for the fundamental transverse electric (TE0) and transverse magnetic (TM0) modes at different thicknesses of the quantum-dot (QD) layer. The index of refraction, $n$, for a quantum-dot film is ~1.7, as indicated by the horizontal dotted line.

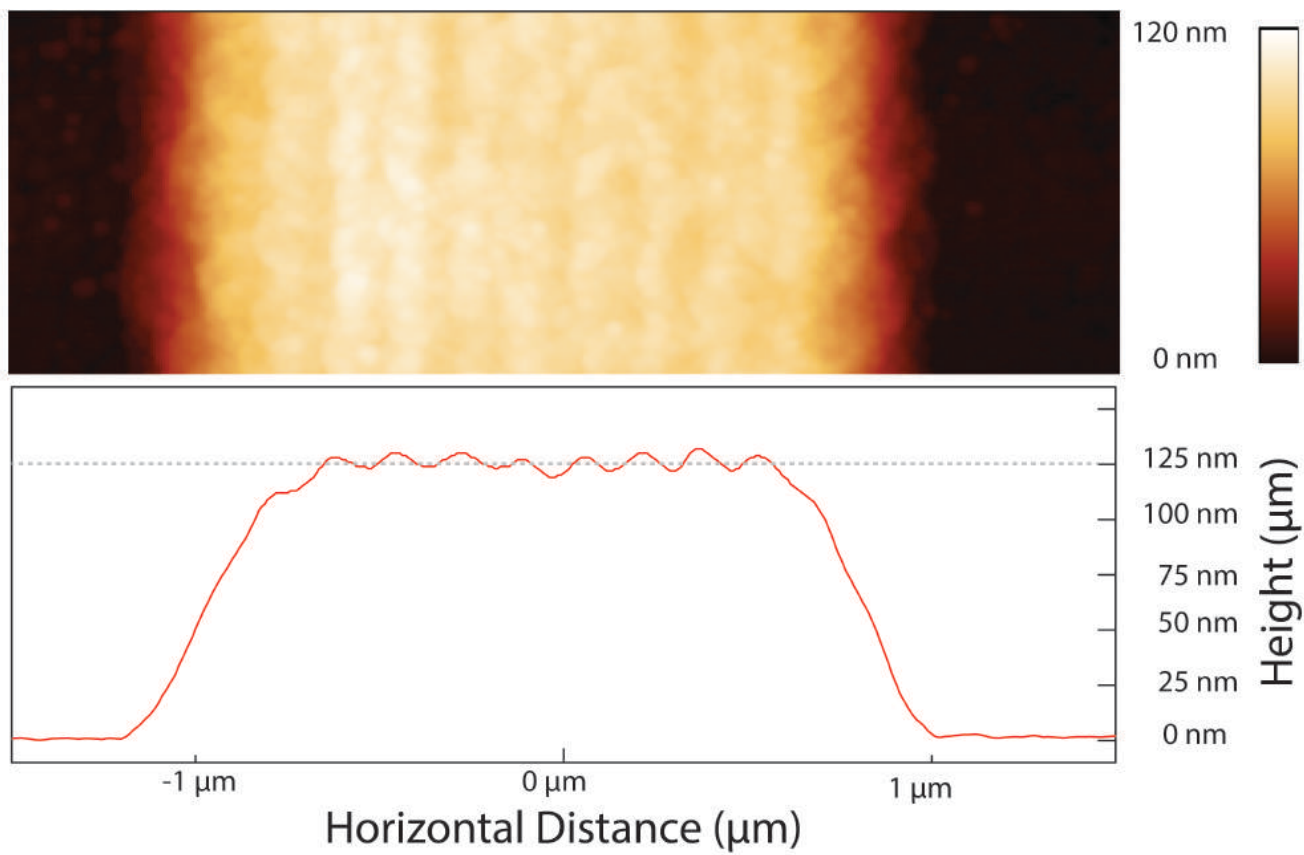

**Supplementary Figure S3 | AFM topographic image of a printed quantum-dot stripe.** The topography of a typical colloidal quantum-dot stripe printed into a spaser cavity reveals that 9 overprints of quantum dots gives a height of ~125 nm.

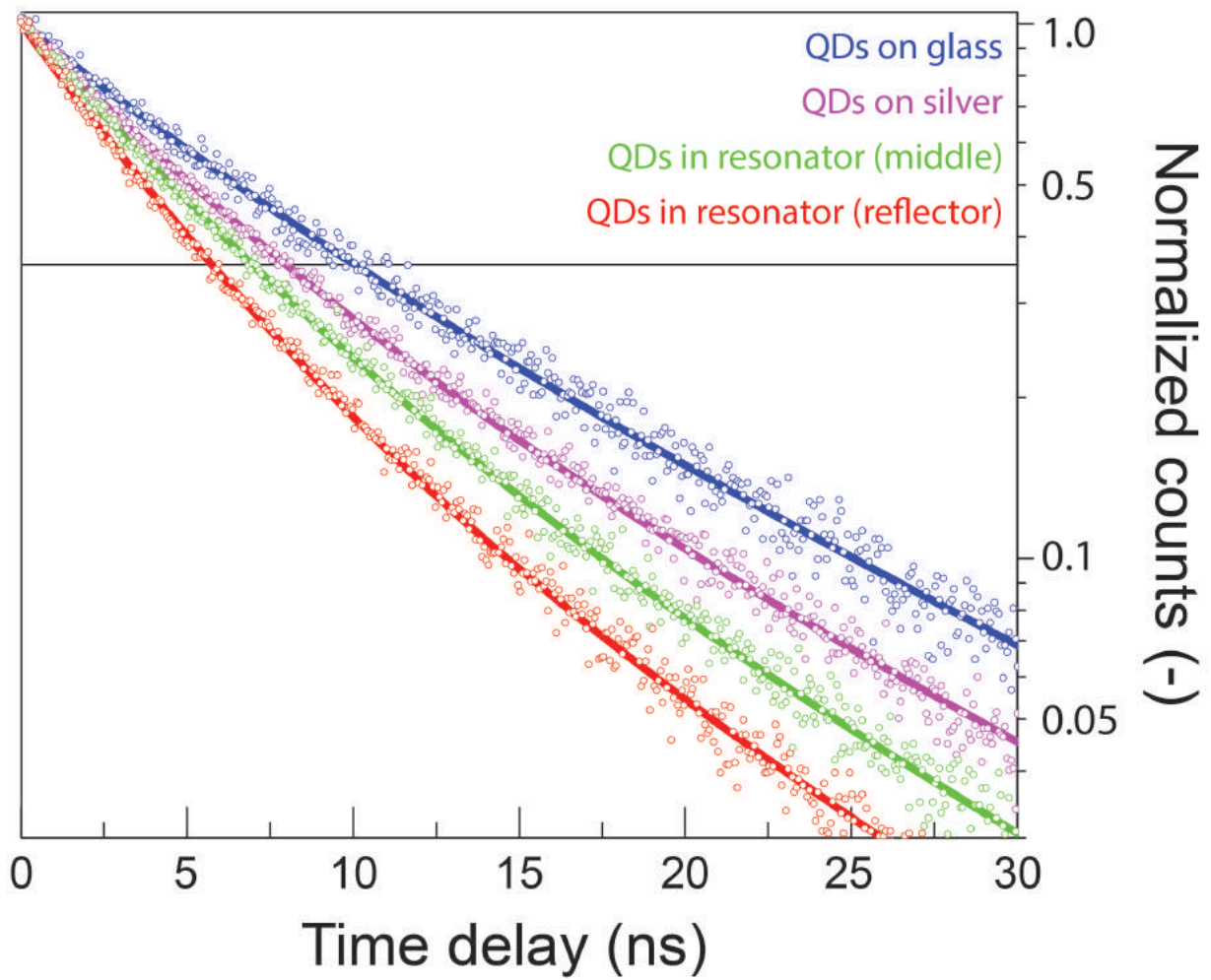

**Supplementary Figure S4 | Lifetime measurements.** The emission decay from quantum dots (QDs) and plasmons scattered from the outer edge of a block reflector measured using time-correlated single-photon counting. A comparison of the decay values at 1/*e* (horizontal line) between the quantum-dot photoluminescence on fused silica and the plasmon scattering from the outer reflector of the cavity indicates a Purcell enhancement of up to 2. See Sections S10 and S11.

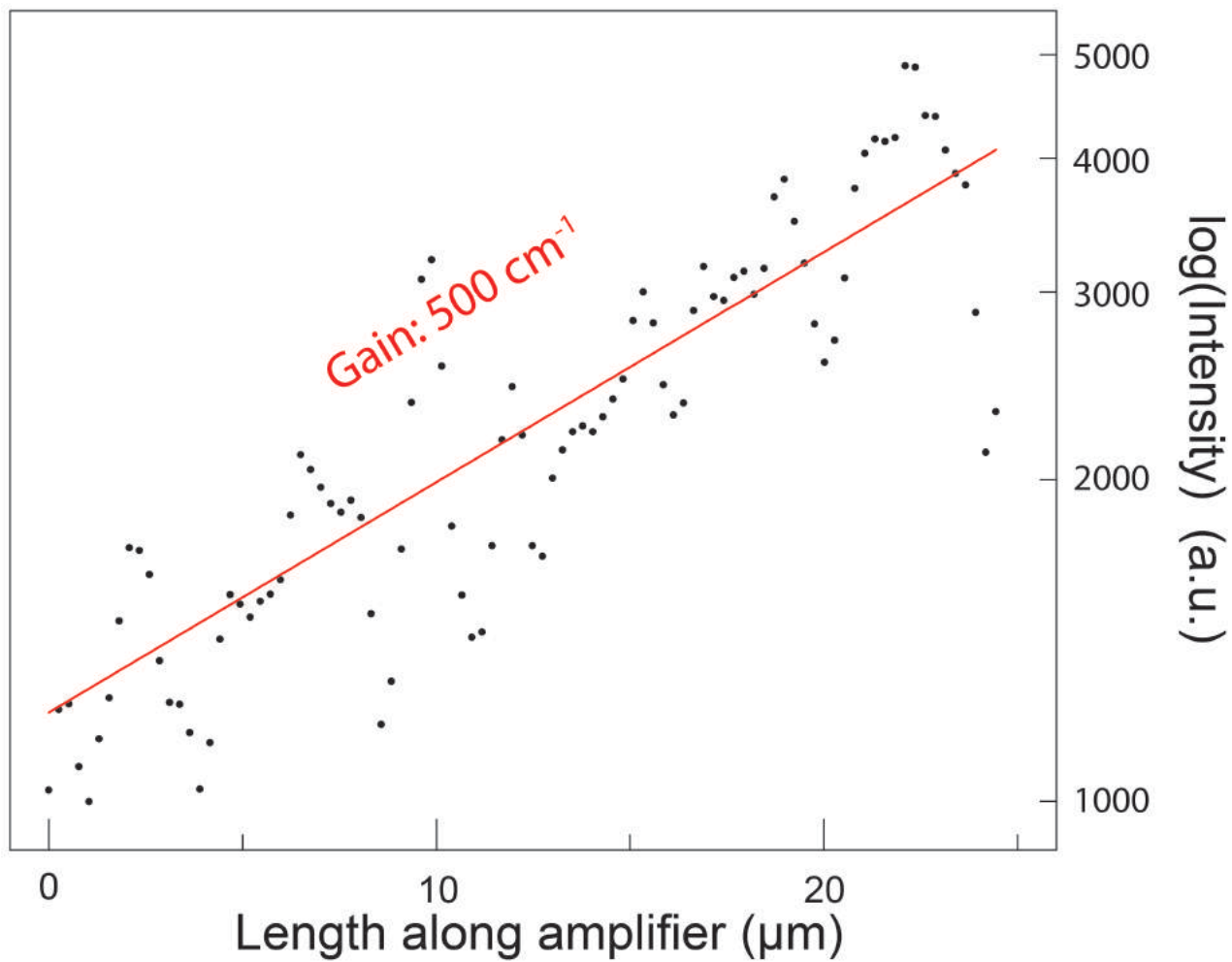

**Supplementary Figure S5 | Measuring amplifier gain.** An exponential fit to the increase in intensity along the length of the bottom edge of the elongated tip structure in Fig. 4 shows an amplification gain of 500 cm$^{-1}$.

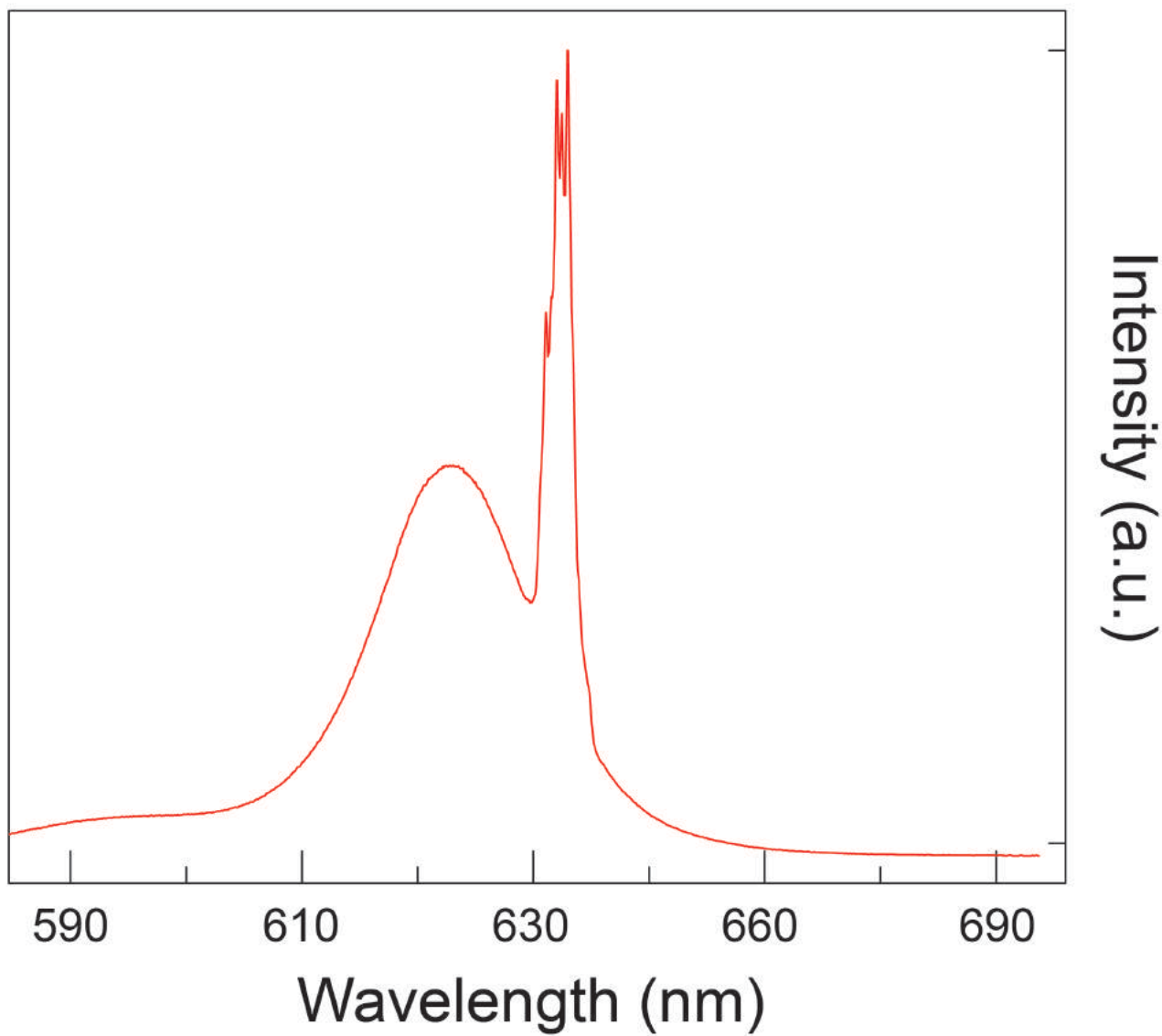

**Supplementary Figure S6 | Amplified spontaneous emission at high excitation intensity.** Excitation of the drop-cast quantum-dot film over a flat portion of the Ag chip shows amplified spontaneous emission at high excitation intensities. Though narrow spectral features can be resolved at the top of the amplified portion of the spectrum, these are unrelated to the spasing peaks observed within the plasmonic cavities.